\documentclass[10pt]{article}
\usepackage[T1]{fontenc}
\usepackage{lmodern}
\usepackage[margin=1in]{geometry}
\usepackage{amsmath,amssymb}
\usepackage{booktabs}
\usepackage{array}
\usepackage{microtype}
\usepackage[hidelinks]{hyperref}
\usepackage{url}

\title{\textbf{Do AI Assistants Fetch Empty Pages More?\\
A Pre-Specified Log Study of ChatGPT-User Fetches,\\
Referred Visits and Index Crawls on a Long-Tail Site}}
\author{Keisuke Watanabe\textsuperscript{*1} and Kazuki Nakayashiki\textsuperscript{\textdagger 1}\\[8pt]\textsuperscript{1}Glasp Inc.}
\date{November 30, 2026}

\begin{document}
\maketitle
\begingroup
\renewcommand{\thefootnote}{\fnsymbol{footnote}}
\footnotetext[1]{Equal contribution; co-corresponding author (\texttt{kei@glasp.co}).}
\footnotetext[2]{Equal contribution; co-corresponding author (\texttt{kazuki@glasp.co}).}
\endgroup

\begin{abstract}
When a person asks an AI assistant a question, the assistant may fetch a web page for that person and may then send the person to it. Site owners see both events in their logs, but to our knowledge prior studies count either crawler requests or referral traffic, not both on the same pages. We report an observational log study on glasp.co, a site of YouTube summary pages, covering tens of thousands of pages that ChatGPT's user-triggered agent (ChatGPT-User) fetched over four weeks (27 October to 23 November 2026). The analysis plan was fixed in a private, time-stamped commit before the outcome window; it is not a public registration and cannot be verified from outside. Three one-sided hypotheses were tested with Holm correction. H1, that empty and missing pages are re-fetched more than article pages, was not supported: the adjusted re-fetch rate ratio for empty versus article pages was 0.84 (95\% CI 0.81 to 0.87). Exploratory checks show this gap is fragile (0.96 (0.94 to 0.97) with all fetches as the outcome; not statistically different from 1 (1.14, 95\% CI 0.98 to 1.32) among pages with four or more pre-period fetches), so the data do not establish the opposite either. H2, that article pages receive more assistant-referred visits than empty pages, was supported: among served visits the ratio ranged from about 1.7 to 2.7 across measures (planned estimate 2.43, 2.22 to 2.66), and it is larger when requests stopped by the site's bot challenge are counted. H3, that more than 30\% of fetched pages had no verified Bingbot or Googlebot crawl in the preceding 28 days, was supported (57.8\%), but this measures how rarely these pages are crawled, not absence from the indexes. Pages were not randomized, so H1 and H2 are associations. The readout ran six days late, and the first three days were collected in one late run that may have lost rows; H2 and H3 should be read as exploratory.
\end{abstract}

\section{Introduction}

AI assistants such as ChatGPT now retrieve web pages while answering a user. These retrievals are made by user-triggered agents that announce themselves in the user-agent string (ChatGPT-User, Claude-User, Perplexity-User) and are distinct from the crawlers that build search indexes (Bingbot, Googlebot, OAI-SearchBot) and from crawlers that collect training data. A site owner who wants to know what these agents mean for the site has two kinds of evidence in the logs: the agent's own fetches, and the human visits that arrive with the assistant as referrer.

Published work has mostly looked at one side. Crawler studies count bots and whether they honor access directives \cite{longpre2024consent,liu2024somesite,kim2025scrapers}. Traffic studies estimate how AI features change visits to sites \cite{khosravi2026aio,zhao2025publishers}. Network operators publish aggregate ratios of crawling to referrals per platform \cite{cloudflare2025crawl}. To our knowledge no open study joins, page by page on one site, the user-triggered fetches of an assistant, the human visits it refers, and the index crawls that preceded the fetch.

This paper reports such a join for one site, glasp.co, whose YouTube summary pages form a large long-tail collection: one page per video, many of which carry a text article written from the video's transcript while others carry none. The design is observational, and its analysis plan was pre-specified in a private commit before the outcome window (Section~\ref{sec:plan}). We asked three questions:
\begin{itemize}
\item[H1] Does the assistant re-fetch pages without an article more often than pages with one? An earlier exploratory observation on the same site suggested that fetch volume runs against content.
\item[H2] Do pages with an article receive more human visits referred by the assistant than pages without one?
\item[H3] Does the assistant fetch many pages that the large search indexes had not crawled recently?
\end{itemize}

The pre-specified hypothesis H1 was not supported. The adjusted re-fetch ratio was below 1, but exploratory checks show that this gap is fragile, and the data do not establish the opposite of H1 either. H2 and H3 were supported as planned. A separately written re-analysis, run after the planned readout, reproduced the planned estimates and added exploratory analyses that qualify the size of H2 and the meaning of H3 (Section~\ref{sec:robust}). We report all three hypotheses at the strength the pre-specified decision rule allows, together with the deviations from the plan, the exploratory analyses labeled as such, and the parts of the measurement we could not verify.

\section{Related work}

\paragraph{Crawler behavior and access control.} Longpre et al.\ audit the web domains behind widely used AI training corpora and document a rapid rise in AI-specific restrictions within one year \cite{longpre2024consent}. Liu et al.\ study whether robots.txt, page-level tags and network-level blocking protect content creators from AI crawlers \cite{liu2024somesite}. Kim et al.\ run controlled robots.txt experiments with self-declared bots and find that compliance depends on the bot and the directive \cite{kim2025scrapers}. Kim et al.\ work from server logs, as we do; Longpre et al.\ audit robots.txt files and terms of service. These studies measure whether bots fetch or are permitted to, not what a fetch leads to.

\paragraph{Crawl-to-referral ratios.} Cloudflare has published network-wide ratios of AI crawling to referrals per platform and a breakdown of AI crawling by purpose \cite{cloudflare2025crawl}. These are aggregates across many sites, with no page-level detail and no variation in page content; they are industry reports, not peer reviewed.

\paragraph{Effects of AI answers on traffic.} Khosravi and Yoganarasimhan use the staggered rollout of Google's AI Overviews and Wikipedia language editions in a difference-in-differences design and estimate a decline in search traffic to English Wikipedia \cite{khosravi2026aio}. Zhao and Berman compare large news publishers that block generative AI crawlers with those that do not and report lower traffic for blockers \cite{zhao2025publishers}. Both measure visits, not assistant fetches. Our own earlier study on the same site used server logs and an untreated part of the domain as a control and concluded that the rise in ChatGPT referrals was suggestive of, but not conclusive for, an effect of content changes \cite{watanabe2026disentangling}.

\paragraph{Content and generative engines.} Generative engine optimization studies change page content and measure visibility in simulated or benchmark engines: Aggarwal et al.\ \cite{aggarwal2024geo} with a fixed set of sources, Bagga et al.\ \cite{bagga2025egeo} with product listings re-ranked by language models, and Ma et al.\ \cite{ma2025goliath}, who observe which content generative search cites and also run a user study on a custom engine. None of these measures real human visits on a live site.

\paragraph{Gap.} We join, per page and per day on one site, verified user-triggered assistant fetches, the human visits referred by the same assistants, the article state of the page, and the index crawls before the fetch.

\section{Setting and data}

\subsection{Site and pages}
The population is the set of pages \texttt{glasp.co/youtube/<video id>} on the apex host. Each page corresponds to one YouTube video. On a given day a page is in one of three states:
\begin{itemize}
\item \emph{article}: a page exists and carries a published article written from the video's transcript;
\item \emph{empty}: a page exists but carries no live article, including pages whose article was withdrawn;
\item \emph{no-doc}: no page record exists; the site then redirects the request to a separate reader view.
\end{itemize}
The plan defines the no-doc group by the redirect. We operationalize it by record existence: a page-day is no-doc when the page's record did not exist at that time, judged by the record's creation time (Section~\ref{sec:dev}). The redirect itself is used only as a check (Section~\ref{sec:robust}).
Articles were not written at random. They were produced in batches on pages selected largely by observed demand, including demand from AI assistants. This selection is the main threat to any comparison between states (Section~\ref{sec:limits}).

\subsection{Request logs}
Requests were taken from the site's content delivery network (Cloudflare), which reports aggregated request counts per minute (per second for assistant fetches) by path, user-agent token, verification flag, network and response status. A daily collector stored these aggregates; the provider keeps them for about a week, so a missed day would have been lost. From 30 October each day was collected on the following day; the first three window days, 27 to 29 October, were collected together in one run on 30 October, up to two days behind that schedule (Section~\ref{sec:complete}). Queries were split until each was below the provider's per-query group limit, and the collector for referred visits failed rather than truncated when a slice stayed at the limit. Absolute page and request counts are withheld; scale is given as orders of magnitude, and results as shares, ratios and intervals. This limits a reader's ability to assess precision and sparse cells independently of the reported intervals.

\paragraph{Fetches (F).} A fetch is a request whose user agent names ChatGPT-User, Claude-User or Perplexity-User and which the network provider marks as a verified bot, on an apex page path, with any status except the host redirect (301). In the window, F consisted almost entirely of ChatGPT-User: verified Claude-User fetches made up under 0.2\% of verified assistant fetches, and no Perplexity-User request was verified.

\paragraph{Referred visits (V).} A referred visit is a request to a page path on the apex host with a referrer or \texttt{utm\_source} value belonging to ChatGPT, Claude or Perplexity, served with status 200, classified by the provider as end-user traffic, not a verified bot, and without a bot-like user agent. Requests that did not reach the page (other statuses, edge subrequests, non-page paths, other hosts) are excluded. V is therefore a count of requests that were served the page, and a lower bound on referred interest. In particular, the site places a bot challenge in front of its pages; a request stopped by the challenge is not served the page and is not counted in V (Section~\ref{sec:limits}).

\paragraph{Index crawls (I).} The plan defines I as index crawls by OAI-SearchBot, Bingbot and Googlebot. H3 as worded uses Bingbot and Googlebot only, so for the H3 test a crawl is a verified Bingbot or Googlebot request to the page path, any status, any host. The share that also counts OAI-SearchBot is reported as a secondary, untested quantity.

\subsection{Page state over time}
\label{sec:state}
Article state per page-day was reconstructed from three timestamps for article creation (the publishing pipeline's log, the record's creation field, and the generation time), taking the earliest, and from the withdrawal timestamp where present. The pipeline log and the creation field agree closely: median difference 0 seconds, more than 99.9\% of pages with both within 10 minutes. The state of a day is the state at 00:00 UTC; a day on which the state changed is dropped from both counts and exposure. One assumption was needed. Pages with a live article but no creation timestamp (about 10\% of fetched pages) are treated as article pages for the whole window. A second group (also about 10\%) consists of withdrawn pages with no recorded article start time; the readout treated them as empty for the whole window and labeled this an assumption, describing them as withdrawn with no withdrawal date. The re-analysis found that this description was wrong: every page in the group carries a withdrawal date, 99.8\% of them before the window began (83\% in July 2026). Empty for the whole window is therefore the correct state for all but about 0.2\% of this group, and only the first group rests on an assumption. Section~\ref{sec:robust} reports an exploratory run that drops both groups.

\subsection{Window and completeness}
\label{sec:complete}
The planned outcome window is 27 October to 23 November 2026 (28 days, UTC). The pre-period used for adjustment is 13 to 26 October. All 28 days have stored files. The days 27 to 29 October were collected together in one run on 30 October, up to two days behind the usual next-day schedule but inside the provider's retention; referred-visit rows on those three days are a little under half the later level (Section~\ref{sec:desc}), and we cannot rule out that this run lost rows. (29 October, collected in that run at the usual one-day lag, is as low as the other two days, so lateness alone does not explain the step.) If rows were lost, V is incomplete on three days, which is the number of missing collection days at which the plan's stop rule makes the readout exploratory. The stop rule as worded (missed days) did not trigger; readers may prefer to treat H2 as exploratory (Section~\ref{sec:plan} extends this to H3).

\section{Pre-specified plan and analysis}

\subsection{Plan}
\label{sec:plan}
The plan was written and frozen by a commit in the company's private repository at 00:22 JST on 26 October 2026 (15:22 UTC on 25 October), before the window began. The analysis plan was thus fixed in a private, time-stamped commit before the outcome window; it is not a public registration and cannot be verified from outside. We therefore call the plan pre-specified, not pre-registered. Appendix~\ref{app:prereg} restates the plan's hypotheses, outcomes and analyses with internal identifiers removed. All three tests are one-sided at $\alpha = 0.05$ with Holm correction across H1 to H3; a hypothesis is supported when its Holm-adjusted $p < 0.05$ and the estimate is in the predicted direction.

The plan left ten choices open (the exact definitions of F and V, the state rules above, the unit of analysis, the meaning of ``re-fetch'', and the H3 lookback); Appendix~\ref{app:choices} lists them. They were fixed in the analysis script, choosing the simplest reading, before the outcome models were run. The script was committed together with the results, after the outcomes had been computed, so the order of the choices relative to the outcome computation rests on the record of the analysis session, not on a timestamp. The 30\% threshold in H3 was set with two earlier values in view, both computed from data before the window: 41.9\% of fetched pages had no Bing, Google or OAI-SearchBot crawl over the lookback then available, and 54.8\% had no Bing crawl in the preceding 14 days.

Given the private plan, the ten choices committed only with the results, the H3 threshold set with two earlier values in view, and the three days collected in one late run (Section~\ref{sec:complete}), the H2 and H3 results should be read as exploratory; we report the planned decisions and these facts together.

\subsection{Models}
\label{sec:models}
H1 and H2 use the same model. The unit is a page $\times$ state segment: all window days a page spent in one state, after dropping transition days. For segment $s$ of page $j$ with outcome $y_s$ over $d_s$ days,
\[
\log \mathbb{E}[y_s] = \log d_s + \beta_0 + \beta_g\,\mathrm{group}_s + \beta_1 \log(1 + \mathrm{pre}_j) + \beta_2 \log(1 + \mathrm{age}_j) + \beta_3\,\mathrm{miss}_j,
\]
with a negative binomial (NB2) likelihood, $\mathrm{Var}(y) = \mu + \alpha\mu^2$. Here $\mathrm{pre}_j$ is the page's fetches in the pre-period, $\mathrm{age}_j$ the video's age in days at the start of the window (from the video's publish time), and $\mathrm{miss}_j$ an indicator for a missing age (under 0.05\% of segments, filled with the median). Standard errors are clustered by page (CR1 sandwich); $\alpha$ is treated as fixed when computing the sandwich, so the standard errors do not include uncertainty in $\alpha$.

\paragraph{H1.} Population: pages with at least one verified ChatGPT-User fetch in the window. Outcome: ChatGPT-User fetches in the window minus the page's first in-window fetch (``re-fetches''). Reference group: article. H1 requires both empty and no-doc pages to exceed article pages, so it is an intersection-union test: each contrast is tested one-sided and the H1 $p$-value is the larger of the two.

\paragraph{H2.} Population: pages with at least one F fetch in the window. Outcome: V. Groups: article versus empty, as in the plan. Pre-period covariate: the page's F in the pre-period. Because verified Claude-User and Perplexity-User fetches were negligible, the H2 population is in effect the ChatGPT-fetched pages, while V includes referrals from all three assistants.

\paragraph{H3.} Population: pages with at least one verified ChatGPT-User fetch in the window. For each, the time of its first in-window fetch; the outcome is whether no Bingbot or Googlebot crawl of the page occurred in the 28 days before. Test: exact one-sided binomial against 0.30; Wilson 95\% interval.

\paragraph{Robustness.} For H1 and H2, a permutation test shuffles group labels between segments within strata of pre-period fetches (0, 1, 2 to 3, 4 to 7, 8 or more), with the log of the exposure-weighted rate ratio (with 0.5 added to each count) as the statistic and 2,000 shuffles; the smallest attainable one-sided $p$ is $1/2001 \approx 0.0005$.

\paragraph{Implementation.} No reference statistics package was available in the analysis environment. The NB2 fit (iteratively reweighted least squares for the coefficients, profile likelihood for $\alpha$), the clustered sandwich, the normal tail, the exact binomial tail and the Wilson interval were implemented in numpy. On simulated clustered count data (200 datasets each at a dense and a sparse base rate), 95\% interval coverage per coefficient was 0.945 to 0.985 and the one-sided size for a null group effect was 0.035. After the readout, a re-implementation written separately (Poisson and NB2 fits with page-clustered errors) reproduced the estimates. It re-loaded the request logs with its own loader; the page records were taken from the readout's stored read and from the same pipeline log files, not read again, so it is independent in code and in the loading of the request logs, not in the page records. Its results: the NB2 rate ratios for the three planned contrasts (0.836, 0.777 and 2.431) and their intervals were identical to three decimals, and the Poisson fits gave 0.83, 0.81 and 2.42. The routines were thus reproduced by an independent implementation, not by an established package. The routines and the simulation are in the ancillary files.

\subsection{Deviations from the plan}
\label{sec:dev}
\begin{enumerate}
\item \emph{Unit and likelihood.} The plan's unit is the page-day, analyzed by a ``negative binomial on page counts''. The fit is instead an NB2 model on page $\times$ state segments with a $\log(\text{days})$ offset. This is a different likelihood from a page-day model: it was chosen to avoid fitting millions of page-day rows. Its estimand is a rate per day spent in a state, and the rate ratios compare those rates between states.
\item \emph{No-doc operationalization.} The plan defines no-doc pages by the redirect to the reader view; the readout uses record existence by creation time instead. Of verified ChatGPT-User requests on page-days labeled no-doc, 66\% received that redirect (Section~\ref{sec:robust}).
\item \emph{Article dating.} The plan dates article state from the publishing pipeline's log and the withdrawal fields. The readout dates the start of an article from the earliest of three sources (the pipeline log, the record's creation field and the generation time), and its end from the withdrawal timestamp.
\item \emph{Read late.} The plan set the readout for 24 November 2026; it ran on 30 November 2026 (UTC). No outcome analysis of this design was run before the readout. However, the site's daily log reports printed the co-fetch share and a Bing lookback share throughout the window, so those two quantities were visible day by day.
\item \emph{H3 lookback.} Log storage began on 6 October, so pages first fetched before 3 November have a lookback of 21 to 27 days rather than 28. A sensitivity analysis restricted to pages with a full 28-day lookback is reported.
\item \emph{Statistical routines} were hand-implemented, as described above, rather than taken from a standard package; they were reproduced afterwards by an independent implementation.
\item \emph{Article-state assumption.} Pages with a live article and no creation timestamp were treated as article pages for the whole window (Section~\ref{sec:state}). An exploratory analysis is reported that drops those pages and also the withdrawn pages with no recorded article start time, which the readout had wrongly counted as a second assumption.
\item \emph{Co-fetch quantity.} The plan words the co-fetch descriptive as a share of fetches in co-fetch clusters; the readout reports the share of clusters that touch two or more pages (Section~\ref{sec:desc}).
\item \emph{Late collection of three days.} The plan assumes daily collection; the days 27 to 29 October were collected together on 30 October (Section~\ref{sec:complete}).
\end{enumerate}

\section{Results}

The H1 and H3 populations comprise tens of thousands of pages; the H2 population is nearly the same set. About two thirds of the H1 segments were in the article state, one third empty, and fewer than 1\% no-doc. Table~\ref{tab:decisions} gives the decisions. Section~\ref{sec:robust} reports exploratory analyses, run after this readout, that bear on how each result should be read; they do not change the decisions.

\begin{table}[t]
\centering
\small
\caption{Planned one-sided tests ($\alpha = 0.05$, Holm across H1 to H3). Holm inputs: the H1 intersection-union $p$-value, the H2 one-sided group test, the H3 binomial test. RR: adjusted rate ratio from the NB2 model.}
\label{tab:decisions}
\begin{tabular}{@{}lllll>{\raggedright\arraybackslash}p{4.4cm}@{}}
\toprule
Hypothesis & Estimate & 95\% interval & One-sided $p$ & Holm $p$ & Decision \\
\midrule
H1 empty vs article & RR 0.84 & 0.81--0.87 & $> 0.99$ & & \\
H1 no-doc vs article & RR 0.78 & 0.62--0.98 & 0.98 & & \\
H1 (intersection-union) & & & $> 0.99$ & $> 0.99$ & Not supported \\
H2 article vs empty & RR 2.43 & 2.22--2.66 & $1.7 \times 10^{-81}$ & $3.5 \times 10^{-81}$ & Supported under the planned test; read as exploratory (rows may have been lost on 27 to 29 Oct) \\
H3 share with no crawl & 0.578 & 0.575--0.582 & $< 10^{-300}$ & $< 10^{-300}$ & Supported under the planned test; read as exploratory (threshold set with earlier values in view) \\
\bottomrule
\end{tabular}
\end{table}

\subsection{H1: empty and missing pages were not re-fetched more}
Adjusted for pre-period fetches and video age, empty pages were re-fetched at 0.84 times the rate of article pages (95\% CI 0.81 to 0.87; $z = -9.6$; $p > 0.99$), and no-doc pages at 0.78 times (0.62 to 0.98; $z = -2.1$; $p = 0.98$). All $p$-values for the planned tests are one-sided, in the pre-specified direction, so an estimate on the other side of the null gives a $p$ near 1. The intersection-union $p$ is $> 0.99$, and the Holm-adjusted $p$ is $> 0.99$. Both contrasts lie in the direction opposite to the hypothesis. Unadjusted, exposure-weighted rate ratios were 0.58 (empty versus article) and 0.65 (no-doc versus article); adjustment narrows the gap but does not reverse it. The stratified permutation test agrees (empty versus article: ratio 0.58, one-sided $p > 0.99$; no-doc versus article: 0.65, $p = 0.82$). The estimated dispersion was $\alpha = 1.21$. Exploratory analyses in Section~\ref{sec:robust} show that the size of the empty-versus-article gap, and in one stratum its sign, depend on analysis choices; we do not read the planned estimate as a finding that article pages are re-fetched more. About a third of the requests on page-days labeled no-doc were not served the defining redirect (Section~\ref{sec:robust}), so the no-doc contrast may be attenuated by misclassification; no-doc is under 1\% of segments, and the H1 decision is determined by the empty versus article contrast.

Covariate estimates, descriptive and not tested: re-fetching is strongly persistent (rate ratio 2.73 per unit of $\log(1+\mathrm{pre})$, 2.68 to 2.80), and older videos were re-fetched slightly less (0.95 per unit of $\log(1+\mathrm{age})$, 0.93 to 0.96).

\subsection{H2: referred visits were higher on article pages}
Among fetched pages, after adjustment for pre-period fetches and video age, article pages received 2.43 times as many assistant-referred visits per day as empty pages (95\% CI 2.22 to 2.66; one-sided lower 95\% bound 2.25; $z = 19.1$; one-sided $p = 1.7 \times 10^{-81}$; Holm $3.5 \times 10^{-81}$); the unadjusted ratio was 2.35. No permutation out of 2,000 reached the unadjusted ratio ($p = 0.0005$, the smallest attainable). Visits were very overdispersed ($\alpha = 13.1$): most pages received none. Covariate estimates, descriptive and not tested: pages with more pre-period fetches (2.19 per unit of $\log(1+\mathrm{pre})$) and older videos (1.40 per unit of $\log(1+\mathrm{age})$) received more visits. The coefficient on the missing-age indicator is degenerate (quasi-separation: the few segments with no known age had no visits) and does not move the group estimate. The size of the ratio is sensitive to how visits are measured (Section~\ref{sec:robust}).

\subsection{H3: ChatGPT-User fetched many pages with no recent verified index crawl}
Of the pages ChatGPT-User fetched in the window, 57.8\% had no verified Bingbot or Googlebot request in the lookback before the first fetch (Wilson 95\% interval 57.5\% to 58.2\%). The exact binomial tail against 0.30 is below the smallest positive double-precision number, reported as $p < 10^{-300}$. Restricted to the about two thirds of pages whose first fetch came on or after 3 November (a full 28-day lookback), the share was 61.4\% (60.9\% to 61.8\%). Truncating the lookback can only raise the share, and only for pages first fetched before 3 November; the full-lookback subset, which truncation does not affect, is well above 0.30, so the decision does not depend on the truncation. (Descriptive, not tested: since a longer lookback can only find more crawls for a given page, the higher share in the subset reflects its different set of pages.) Secondary, untested shares: no Bingbot, Googlebot or OAI-SearchBot crawl, 46.7\% (51.3\% with full lookback); no Bingbot crawl alone, 65.2\% (69.2\%). What this share does and does not measure is taken up in Section~\ref{sec:robust}.

\subsection{Pre-specified descriptives}
\label{sec:desc}

\paragraph{Fetch-to-visit ratio.} As a descriptive site-level ratio, not a per-fetch conversion rate, there were 0.81 ChatGPT-referred visits (V) per verified ChatGPT-User fetch over the window (1.24 fetches per visit). V here counts visits on every page of this type, including pages ChatGPT-User did not fetch in the window. Two cautions apply to it (exploratory, from the re-analysis). First, V counts only visitors who passed the site's bot challenge. Second, the daily number of referred-visit rows in the store roughly doubles from 30 October: mean daily V from 30 October on is 2.3 times that of 27 to 29 October, while verified fetches do not rise over the same days (they decline gradually over the window). The three low days are the days that were collected late in one batch (Section~\ref{sec:complete}). We have not determined whether the step is a loss of rows in that late collection or a real change in traffic. In an exploratory check, the unadjusted H2 ratio is 2.16 on 27 to 29 October, 2.36 from 30 October and 2.35 over the whole window, so the comparison between groups is similar in the two periods. The site-level figure is not: it is 0.31 over the first three days and 0.88 from 30 October. For Claude the ratio was 0.21, on a verified base too small to interpret; for Perplexity it is undefined because no Perplexity-User fetch was verified.

\paragraph{Co-fetch clusters.} Verified ChatGPT-User fetches were grouped into clusters of requests from the same network with gaps of at most 3 seconds. The share of clusters touching two or more pages was 0.243, against 0.162 in a placebo in which each request's second was redrawn uniformly within its hour. The plan words this item as a share of fetches; we report the share of clusters, which is the quantity behind the plan's own exploratory values (0.297 and 0.223). A gap of about 0.07 to 0.08 persisted at a lower level.

\paragraph{Unverified self-declared bots.} Table~\ref{tab:unverified} gives, per user-agent token, the share of requests that the network provider did not verify, and the part of all requests that was unverified and came from networks outside those listed for that token in the collector (the vendor's own networks, or the cloud networks the vendor is known to use). For Googlebot, Applebot and ChatGPT-User most unverified traffic came from the networks listed for the vendor. The re-analysis (exploratory) looked at how these requests relate to the verified ones. For ChatGPT-User, the unverified requests from the network listed for it and the verified requests are almost one to one: their daily ratio stayed between 0.96 and 1.17 over the window (1.01 overall), and the two counts were equal on about two thirds (0.668) of the page-days on which a fetched page had either. For Googlebot, over the stored period and on all pages of this type, 85.3\% of the unverified log rows from Google's own network have a verified Googlebot row for the same page within one minute. Bingbot has almost no unverified requests, so no such pairing can be assessed for it; Applebot was not examined. The unverified half of ChatGPT-User and Googlebot requests is therefore probably a second logged request belonging to the same fetch or crawl, not separate activity; its cause is not known. For Claude-User most (about 87\%) of the unverified traffic came from the cloud networks that vendor uses, and we cannot rule out other clients on the same clouds. Perplexity-User was never verified, and almost all of it came from networks outside those listed for it; we cannot tell whether these are Perplexity's agents running elsewhere or other clients using the name.

\begin{table}[t]
\centering
\small
\caption{Share of requests naming each bot that the network provider did not verify, over the window. ``Outside listed networks'': the part of all requests that was unverified and came from networks outside those listed per token. The listed networks are the vendor's own networks and the cloud providers the vendor is known to use, as configured in the collector. n/a: no network was listed for DuckAssistBot, so all of its unverified traffic falls outside by construction. All tokens the collector tracked are shown except Amzn-SearchBot and meta-webindexer, which almost no requests named.}
\label{tab:unverified}
\begin{tabular}{@{}lrr@{}}
\toprule
User-agent token & Unverified & Unverified, outside listed networks \\
\midrule
ChatGPT-User & 0.506 & 0.010 \\
OAI-SearchBot & 0.002 & 0.002 \\
GPTBot & 0.004 & 0.004 \\
Claude-User & 0.998 & 0.131 \\
Claude-SearchBot & 0.020 & 0.000 \\
ClaudeBot & 0.021 & 0.007 \\
Perplexity-User & 1.000 & 0.991 \\
PerplexityBot & 0.916 & 0.054 \\
Google-Agent & 0.041 & 0.000 \\
Googlebot & 0.453 & 0.014 \\
bingbot & 0.003 & 0.003 \\
Applebot & 0.436 & 0.001 \\
MistralAI-User & 0.095 & 0.000 \\
DuckAssistBot & 0.426 & n/a \\
\bottomrule
\end{tabular}
\end{table}

\section{Exploratory re-analysis and data checks}
\label{sec:robust}

Every analysis and data check in this section is exploratory: none was pre-specified, and all were made after the results of the planned tests were known. The first analysis and the two data checks at the end were run with the readout's own script and its logs. The other three analyses come mostly from a re-analysis with separately written code, which re-loaded the request logs, took the page records from the readout's stored read (Section~\ref{sec:models}), reproduced the planned estimates, and then looked for ways in which the three results could be read too strongly; a few of the figures were added in the revision and review that followed it. These analyses were chosen with the results in view and are not corrected for multiplicity. They do not change the three planned decisions; they change what the decisions can be taken to mean.

\paragraph{Undated pages (exploratory).} This analysis drops two groups, about 20\% of fetched pages together: pages with a live article and no creation timestamp, whose state rests on an assumption, and withdrawn pages with no recorded article start time, which the readout also treated as resting on an assumption although their state is in fact known (Section~\ref{sec:state}). H1: empty versus article 0.88 (0.85 to 0.92), no-doc versus article 0.77 (0.61 to 0.97). H2: 2.71 (2.43 to 3.01), $p = 2.3 \times 10^{-73}$; permutation ratio 2.49, $p = 0.0005$. H3 does not use article state. The decisions do not depend on these two groups.

\paragraph{H1: the gap is fragile (exploratory re-analysis).} The planned outcome is re-fetches, that is, fetches minus the page's first in-window fetch. With all window fetches as the outcome and the same model, the adjusted empty-to-article ratio is 0.96 (0.94 to 0.97); unadjusted, without subtracting the first fetch, it is 0.80, against 0.58 for re-fetches. Fitting the planned model for empty versus article pages within strata of pre-period fetches gives 0.812 (0.777 to 0.848) for pages with no pre-period fetch, 0.878 (0.817 to 0.944) for pages with 1 to 3, and 1.136 (0.976 to 1.323) for pages with 4 or more. Article pages had about three times the pre-period fetches of empty pages on average, and the pre-period count is zero for most pages in both groups (71\% of article segments and 86\% of empty segments), so for most pages the covariate carries no information about prior demand. The remaining gap is therefore consistent with selection on demand that the covariate does not remove; we cannot show that this is its cause. The pre-specified hypothesis was not supported, and the data do not establish the opposite either.

\paragraph{H2: the size depends on the measure (exploratory re-analysis).} Visits are concentrated on few pages: the share of segments with any referred visit was 0.148 for article and 0.086 for empty segments. With the outcome reduced to whether a segment had any visit (Poisson model, same covariates and offset, page-clustered errors), the adjusted article-to-empty ratio of the share of segments with any visit is 1.74 (1.65 to 1.82). Unadjusted variants: capping V at 10 per segment gives 2.11 (2.35 uncapped); dropping the top 1\% of pages by V gives 1.78; visits per fetch in place of visits per page-day gives 1.87; by period, 2.16 on the three days collected late and 2.36 afterwards (Section~\ref{sec:desc}). Article pages received more on every measure we tried. Among served visits the ratio ranged from about 1.7 to 2.7 (the planned estimate is 2.43; the run without undated pages, above, gave 2.71), and it is larger, not smaller, when challenged requests are counted (Section~\ref{sec:limits}). Two further limits on what V measures, the site's bot challenge and the share of referred visits that fall outside the H2 population, are described in Section~\ref{sec:limits}.

\paragraph{H3: what the share measures (exploratory re-analysis).} On the full-lookback subset, the share of pages with no verified Bingbot or Googlebot crawl before the first fetch falls steadily as the lookback lengthens: 0.902 at 7 days, 0.823 at 14, 0.726 at 21 and 0.614 at 28, with no plateau. Over the whole stored period (6 October to 23 November), counting crawls before or after the fetch, the share of fetched pages with no verified Bingbot or Googlebot crawl at all is 0.413, and it was still falling when the period ended (0.434 one week earlier). The share at the planned lookback differs by the page's state at its first fetch: 0.485 for article pages and 0.772 for empty pages. Counting unverified Googlebot requests from Google's own network and unverified bingbot requests from Microsoft's as crawls gives 0.571 in place of 0.578. H3 as planned is supported. But the quantity measures how rarely the indexes re-crawl these pages within 28 days, not whether the pages are absent from the indexes: a page that was not crawled in 28 days can still be indexed, and a longer log would probably have found a crawl for more of these pages.

\paragraph{No-doc label against the server's response.} Of verified ChatGPT-User requests on page-days labeled no-doc, 66\% received the redirect to the reader view and 29\% a normal page; the rest were closed by the client. No redirect to the reader view (307) occurred on page-days labeled article or empty. The label is mostly right; some no-doc page-days are probably records created and deleted, or created earlier than recorded. This was not checked further.

\paragraph{Collection.} No collection slice approached the per-query group limit; over all 28 days, the largest number of groups any single bot produced in one minute was far below the limit. Some days were collected twice; the copies differ by less than 0.01\% of rows, and one copy is used. From 30 October every window day was collected on the following day; 27 to 29 October were collected together on 30 October (Section~\ref{sec:complete}). The step in daily referred-visit rows from 30 October coincides with that late batch; whether rows were lost was not determined (Section~\ref{sec:desc}).

\section{Discussion and limitations}
\label{sec:limits}

\paragraph{What the results say.} On this site, in this window, pages that carry an article received more assistant-referred human visits than empty pages that were also being fetched. They received more on every measure we tried; among served visits the ratio ranged from about 1.7 to 2.7 (planned estimate 2.43, 2.22 to 2.66), and it is larger, not smaller, when challenged requests are counted. The pre-specified expectation that empty and missing pages are re-fetched more did not hold. The adjusted re-fetch ratio was below 1 overall, but this is fragile (Section~\ref{sec:robust}): it nearly closes when all fetches are counted, and among pages with four or more pre-period fetches it is above 1 with an interval that includes 1. We do not take it as evidence that article pages are re-fetched more. More than half of the pages the agent fetched had no verified Bingbot or Googlebot crawl in the preceding three to four weeks. That share measures how rarely such long-tail pages are crawled; it falls steadily with a longer lookback and is lower for article pages than for empty ones. This study cannot say whether those pages were in a search index when the assistant fetched them.

\paragraph{Selection.} Articles were written on pages chosen largely by demand, including assistant demand. Adjustment for each page's pre-period fetches and for video age may remove part of this, but not all: a page chosen for an article because it was in demand may stay in demand for reasons unrelated to the article, and the pre-period count is zero for most pages, so it cannot rank them. The empty group is selected as well. Most withdrawn pages were withdrawn in July 2026 under a rule that selected articles flagged as having no search visibility and with no recorded AI-assistant requests in the preceding weeks (confirmed on a random sample of the page records, all of which carry that rule's label). The empty group is therefore partly selected on low search and assistant activity at the time, which bears on H1 and on the H3 share for empty pages as well as on H2. The withdrawal rule likely biases H1 toward fewer re-fetches on empty pages and raises the H3 no-crawl share for empty pages. (The exploratory run that left out most of the withdrawn pages, together with the undated live pages, gave a larger H2 ratio, 2.71, not a smaller one.) All H1 and H2 comparisons are also conditional on the page being fetched in the window, and the pre-period adjustment is zero for most pages, so selection on what gets fetched is only partly addressed. H1 and H2 are associations. A randomized assignment of articles to pages already fetched by assistants would be needed to estimate the effect of writing an article.

\paragraph{Instrument.} F depends on the network provider's verification. Nearly all Claude-User and Perplexity-User requests were unverified, so the Claude and Perplexity parts of the plan could not be measured. Half of the requests naming ChatGPT-User were unverified too, but these appear to be paired with verified requests (Section~\ref{sec:desc}); their cause is not known, and we cannot confirm that each pair is one fetch. The same holds for H3: a crawl counts only when verified, and 45\% of requests naming Googlebot were unverified, most of them from Google's own networks and most of those paired with a verified request; counting them as crawls moves the H3 share from 0.578 to 0.571. Agents that fetch without announcing themselves are invisible.

V counts only requests that were served the page with an assistant referrer or tag. A large part of assistant-referred requests is stopped by the site's bot challenge before the page is served, and that part is larger on article pages than on empty pages. V is therefore a count of visitors who passed the challenge. In an exploratory check the H2 ratio is larger, not smaller, when challenged requests are counted, so the direction holds but the size of the ratio depends on how they are treated. Visits whose referrer was stripped are missed. Attribution of V relies on the referrer or \texttt{utm\_source} value, which a user or a third party can set; this was not checked. The H2 population (pages with a verified fetch in the window) receives only about one third of all assistant-referred visits to pages of this type; the rest land on pages with no verified fetch in the window, whose article state was not read. V on 27 to 29 October, the three days collected late in one batch, is a little under half the later level, and we cannot rule out that rows were lost; if they were, V is incomplete on three days, the number at which the plan's stop rule would make the readout exploratory (Section~\ref{sec:complete}). The unadjusted H2 ratio is similar on those days (2.16) and afterwards (2.36). Pages on the same query may interfere with each other, which a page-level model ignores.

\paragraph{Analysis.} The readout ran six days late, and the first three window days were collected in one late run. The ten interpretation choices were committed only together with the results. The H3 threshold was set with two earlier values, from data before the window, in view. The unit, the no-doc rule and the article dating differ from the plan's wording (Section~\ref{sec:dev}). The statistical routines were checked by simulation and reproduced by a separately written implementation that shared the readout's page records, not by a reference package. Article states for about a tenth of fetched pages (live articles with no creation timestamp) rest on an assumption, although the decisions do not change without them. The exploratory analyses were chosen after the results were known. A record's history of earlier withdrawals and restorations is not visible in its current fields, and a record deleted before the readout reads as no-doc for the whole window.

\paragraph{Conduct.} The collector, the analysis, the manuscript and the reference check were produced with heavy AI assistance (see Statements). The references were not read in full by the authors, and the positioning against prior work should be read as provisional. The study was run by the site's operator on its own data, under a plan that was private, with no external audit.

\paragraph{Generality.} This is one site, with one content type, in one four-week window, dominated by one assistant. The results could differ on other sites or as assistants change how they retrieve pages.

\section{Conclusion}
In a log study with a privately pre-specified plan on a long-tail site, the pre-specified hypothesis that ChatGPT's user-triggered agent re-fetches empty or missing pages more than article pages was not supported; the data do not establish the opposite either. Article pages received more assistant-referred visits than empty pages that were also fetched, on every measure we tried: among served visits the ratio ranged from about 1.7 to 2.7 (planned estimate 2.43, 2.22 to 2.66), and it is larger, not smaller, when requests stopped by the site's bot challenge are counted. Most fetched pages had no verified Bingbot or Googlebot crawl in the preceding 28 days, which shows how rarely such pages are crawled, not that they are missing from the indexes. H1 and H2 are associations on non-randomized pages, and given the private plan, the late collection run and the H3 threshold, the H2 and H3 results should be read as exploratory. A randomized holdout of article creation among pages that assistants already fetch is the natural next step.

\section*{Statements}

\paragraph{AI use.} Claude (Anthropic, Claude Opus 5.5) wrote the log collector used to gather the data; one change to the collector (failing rather than truncating at the query limit) followed an AI code review by Codex (OpenAI). The pre-specified analysis plan was also drafted by Claude Opus 5.5. Claude Opus 5.5 also wrote the analysis script, ran the planned readout, and drafted this manuscript and its tables. The reported numbers were checked against the analysis output by separate Claude sessions. The independent re-analyses were run by Claude Opus 5.5 and Claude Fable 5.2 (Anthropic) sessions: they re-loaded the request logs with separately written code, took the page records from the readout's stored read, reproduced the planned estimates and produced most of the exploratory analyses in Section~\ref{sec:robust}. Manuscript reviews were carried out by Claude Fable 5.2 sessions and, on 30 November 2026 (UTC), in two passes by an OpenAI model (GPT-5) acting as a referee. Its points on framing were applied; its requests for absolute sample sizes, for a figure for the share of challenged requests, for a commitment to a replication package or audit log, and for further sensitivity analyses were not. The references were checked against their abstract pages by an AI session; they were not read in full by the authors. No review by a person other than the authors was carried out. The authors approved the submission. No AI system is an author.

\paragraph{Competing interests.} Both authors are co-founders of Glasp Inc., which operates glasp.co, the site studied. Glasp benefits from traffic to these pages and from knowledge of how AI assistants treat them. The data are Glasp's own logs and the study was carried out by Glasp.

\paragraph{Funding.} This study received no specific grant; it was carried out with Glasp Inc.'s own resources.

\paragraph{Data and code availability.} The request logs are the site's own access logs and are not released. The statistical routines used for the planned tests (the NB2 fit, the clustered sandwich, the binomial tail, the Wilson interval, the Holm step and the permutation test), the coverage simulation and a demonstration on synthetic data are provided as ancillary files with this version; they contain no site data and no data loaders. Whether page-level aggregates and the remaining analysis code will be released has not been decided.

\paragraph{License.} This work is licensed under CC BY 4.0.

\appendix
\section{The pre-specified plan, restated}
\label{app:prereg}

Restated from the frozen plan of 26 October 2026, with internal identifiers removed. The hypotheses are quoted, without the plan's labels and parentheticals; the other parts are paraphrased.

\paragraph{Population and unit.} Page paths of the form above on the apex host; the plan's unit is the page-day. (The readout fits page $\times$ state segments with a log-days offset instead; see Section~\ref{sec:dev}.)

\paragraph{Window.} In the plan's words, ``Confirmatory'': 27 October to 23 November 2026 UTC (4 weeks). Data before 27 October is ``exploratory''.

\paragraph{Groups.} Article page (a page record with a published article that day), empty page (record, no article), no-doc page (redirect to the reader view).

\paragraph{Outcomes.} F: user-triggered assistant fetches per page-day by ChatGPT-User, Claude-User and Perplexity-User, verified by the network provider. V: human visits with that assistant's referrer or \texttt{utm\_source} that reached the page (status 200, end-user traffic, served the page itself). I: index crawls (OAI-SearchBot, Bingbot, Googlebot) in the 28 days before a fetch.

\paragraph{Hypotheses} (one-sided, $\alpha = 0.05$ each, Holm across H1 to H3).
\begin{itemize}
\item[H1] ``Among pages with at least one ChatGPT-User fetch in the window, re-fetches per page are higher for empty and no-doc pages than for article pages, after adjusting for fetches in the 2 weeks before 10-27.''
\item[H2] ``V per fetched page is higher for article pages than for empty pages.''
\item[H3] ``More than 30\% of pages ChatGPT-User fetches had no Bingbot or Googlebot crawl in the prior 28 days.''
\end{itemize}
Descriptive, no test: fetch-to-visit ratio per assistant; share of fetches in co-fetch clusters against the placebo; share of traffic that calls itself an AI bot and is not verified.

\paragraph{Analysis.} H1, H2: negative binomial on page counts with group, pre-period fetches and video age; page-clustered errors; a permutation test as the robustness check. H3: binomial with a Wilson interval. Pages are not randomized, so H1 and H2 are associations; in the plan's words, ``a group difference that vanishes with pre-period adjustment is reported as such.''

\paragraph{Labels.} The plan labeled H3 ``assistants fetch outside the indexes''; the data do not support that reading (Section~\ref{sec:robust}). It labeled H1 ``fetch runs against content''; the data do not support that reading either.

\paragraph{Known threats named in the plan.} Selection (articles were written on pages chosen by demand); bots that fetch without a user agent; the site's bot challenge on human visits, which stops some before they reach the page; interference between pages on the same query. Stop rule: if the collector misses 3 or more days in the window, the readout is exploratory.

\section{The ten analysis choices}
\label{app:choices}

Choices the plan left open, as fixed in the analysis script; by the analysis session's record, before the outcome models were run (Section~\ref{sec:plan}). Internal identifiers are removed.

\begin{enumerate}
\item[C1] F counts requests the network provider marks as verified bots, any status except 301 (the redirect from the www host to the apex host, so the count is apex only), on page paths with an 11-character video id and an optional trailing slash.
\item[C2] The article state of a page-day is the state at 00:00 UTC; a day on which the state changed (article published, article withdrawn, record created) is dropped from both counts and exposure, so no day mixes two states.
\item[C3] An article lasts from the earliest of three timestamps (the publishing pipeline's log, read in the pipeline's local time zone and converted to UTC; the record's creation field; the generation time) to the withdrawal timestamp if the article is withdrawn now. A record with a live article and no timestamp counts as an article for the whole window; a withdrawn record with no withdrawal timestamp counts as empty for the whole window. (As applied, this last rule covered withdrawn records with no recorded article start time; all of them had a withdrawal timestamp, see Section~\ref{sec:state}.) A record exists from its creation time; before that the page is no-doc.
\item[C4] An empty page is one whose record exists and that has no live article that day, including records whose article was withdrawn.
\item[C5] The unit for H1 and H2 is a page $\times$ state segment (all window days of a page in one state, after C2), with the outcome summed over those days and an offset of log(days in that state). NB2 with log link; covariates: group, $\log(1+\text{pre-period fetches})$, $\log(1+\text{video age in days at 27 October})$ with a missing-age indicator (age filled with the median). Errors clustered by page; $\alpha$ treated as fixed in the sandwich.
\item[C6] H1 ``re-fetches'' are verified ChatGPT-User fetches in the window minus the page's first in-window fetch. Population: pages with at least one such fetch. Pre-period covariate: the page's verified ChatGPT-User fetches from 13 to 26 October. H1 asks that both empty and no-doc pages exceed article pages: each contrast is tested one-sided and the H1 $p$-value is the larger of the two (intersection-union test).
\item[C7] H2 population: pages with at least one verified fetch by ChatGPT-User, Claude-User or Perplexity-User in the window; article versus empty segments only (no-doc pages redirect, so V is zero there by construction). Note: that reason no longer holds exactly under the record-existence rule of C3, since some no-doc page-days were served a normal page (Section~\ref{sec:robust}); the exclusion stands because the plan's H2 compares article with empty pages only. V: requests on the apex host with status 200, classified as end-user traffic, without a bot-like user agent, not a verified bot, and with a referrer or \texttt{utm\_source} of one of the three assistants. Pre-period covariate: the page's F from 13 to 26 October.
\item[C8] H3: pages ChatGPT-User (verified) fetched in the window; a crawl is a verified Bingbot or Googlebot request to the page (any status, any host); the lookback is 28 days before the page's first in-window fetch, cut at the first stored day, 6 October (so 21 to 27 days for first fetches before 3 November, the full 28 days on or after). Test: exact one-sided binomial against 0.30; Wilson 95\% interval. Sensitivity: pages whose first fetch is on or after 3 November.
\item[C9] Permutation check: group labels shuffled among segments within strata of pre-period fetches (0, 1, 2 to 3, 4 to 7, 8 or more); statistic: the log of the exposure-weighted rate ratio (with 0.5 added to each count); 2,000 shuffles; one-sided.
\item[C10] Holm: the three one-sided $p$-values (H1 as in C6) are adjusted together; a hypothesis is supported when its Holm-adjusted $p < 0.05$ and the estimate is in the predicted direction.
\end{enumerate}

\end{document}